\documentclass[12pt]{iopart} 
\usepackage{graphicx}   
\begin{document} 
 
\title{Quantifying the taxonomic diversity in real species communities} 
 
\author{Cecile Caretta Cartozo} 
\address{Laboratoire de Biophysique Statistique, ITP-FSB, Ecole Polytechnique F\'ed\'erale de Lausanne,  
CH-1015 Lausanne, SWITZERLAND} 
\ead{cecile.carettacartozo@epfl.ch} 
\author{Diego Garlaschelli} 
\address{INFM UdR Siena, Department of Physics, University of Siena, Via Roma 56, 53100  Siena, ITALY} 
\ead{garlaschelli@unisi.it} 
\author{Carlo Ricotta} 
\address{Department of Plant Biology, University of Rome ``La Sapienza'', Piazzale Aldo Moro 5, 
00185 Rome, ITALY} 
\ead{carlo.ricotta@uniroma1.it} 
\author{Marc Barth\'elemy} 
\address{School of Informatics and Biocomplexity Center, Indiana University, 
Eigenmann Hall, 1900 East Tenth Street, Bloomington, IN 47406, USA} 
\ead{mbarthel@indiana.edu} 
\author{Guido Caldarelli} 
\address{INFM-CNR Centro SMC Department of Physics, University of Rome ``La Sapienza'',  
Piazzale Aldo Moro 5, 00185 Rome, ITALY} 
\ead{guido.caldarelli@roma1.infn.it} 
 
\begin{abstract} 
We analyze several {\em florae} (collections of plant species populating specific areas) in different  
geographic and climatic regions. For every list of species we produce a taxonomic classification tree  
and we consider its statistical properties. We find that regardless of the geographical location,  
the climate and the environment all species collections have universal statistical properties that we  
show to be also robust in time. We then compare observed data sets with simulated communities obtained  
by randomly sampling a large pool of species from all over the world. We find differences in the behavior  
of the statistical properties of the corresponding taxonomic trees. Our results suggest that it is  
possible to distinguish quantitatively real species assemblages from random collections and thus demonstrate 
the existence of correlations between species.  
\end{abstract} 
 
\maketitle 
 
\section{Introduction} 
The search for patterns in the species composition across ecological communities and for the  
processes that cause these patterns is a central goal of ecology. The starting point is the general  
assumption that an ecosystem is not a simple random collection of species. A great variety of evolution  
models suggest that an ecosystem is shaped by a number of evolutionary mechanisms such as speciation,  
competition and selection that differentiate it from a random system. Several approaches try to show  
evidences for these differences, even if slight. Our contribution to these efforts focuses on the study  
of the structure of the taxonomic classification of plants species. Though modern systematic biologists  
consider it more reliable and less arbitrary to describe relationships between species via phylogenetic  
trees \cite{WebbandPitman2002,Hennig1966} (hierarchical structures following the steps of evolution from 
ancestral species to modern ones),  
the usual taxonomic classification is still widely used. Introduced in the XVIII century by the  
Swedish naturalist Carl von Linn\'e, taxonomy is based on morphological and physiological observations and  
it groups all species in a set of different hierarchical levels very much similar to a genealogical tree. Thus, with the 
aim of comparing our results to the literature in the field we restrict our analysis to the traditional Linnean 
taxonomy. Nonetheless, due to the positive relationship between 
the  phylogenetic closeness of two taxa and their morphological similarity,\cite{MooersandHeard1997}   
the taxonomic tree of a particular flora still contains information on the 
processes that shape it.  
Any classification group in the taxonomic tree is generally called a {\it taxon}. The distribution of the number 
of subtaxa per taxon at one specific taxonomic level has been widely studied  
starting with the work of Willis\cite{Willis1922} and Yule \cite{Yule1924}. 
Willis observed that the  
distribution of the number $n_g$ of genera containing a number $n_s$ of species 
in the set of all flowering  
plants is a power-law of the kind $n_g \propto n_s^{-\gamma}$ with an exponent $\gamma=-1.5$.  
Two years later Yule proposed a branching process model to explain this distribution.  
Since then, the shape of taxonomic abundance distributions has been the object of a large number of  
studies (For general reviews in the field see Raven \cite{Raven1976}   
Burlando \cite{Burlando1990,Burlando1993} and references therein. For details on the connection between  
phylogenesis and community ecology see Webb et al \cite{WebbAckerlyetal2002}).  
Burlando \cite{Burlando1993} extended these results to other pairs of taxonomic levels and observed  
similar behaviors for the distribution of the number $n_t$ of taxa with $n_{st}$ subtaxa. 
Models have been proposed in order to reproduce this behavior  
\cite{Enquistetal2001,Enquistetal2002,Webb2000, Epsteinetal1989}. 

In this paper, we study the topological properties of taxonomic trees with the specific aim of  
shading some light on the possible qualitative and quantitative differences between real ecosystems  
and random species assemblages. Two main points distinguish our approach from those already present  
in the literature. First of all, instead of considering single taxonomic levels (as in the case 
of Willis \cite{Willis1922} and Burlando \cite{Burlando1993}), we present results  
obtained from the study of the statistical properties of the entire taxonomic tree as an all.  
Secondly, unlike previous studies that focused on specific taxonomic groups, like Fabaceae, Poaceae or
Brassicaceae, we analyze florae resulting from plants collected in the same ecosystem or at least in a specific
geographic and climatic area. These plants do not necessarily belong to the same taxonomic group. 
They represent species that actually coexist, sharing the same environment and the same 
ecological conditions.   

\section{Data and Methods}
\subsection{Taxonomic data sets and graph representation}
We studied $22$ different florae with different geographical and climatic properties from all around the world. 
The size of the single data sets spreads from about $100$ to $40,000$ species. 
All species considered in the different florae belong to the phylum of vascular plants 
(earthly plants with 
a lymphatic system). Most data sets have been kindly made available by the Department of 
Plant Ecology of 
the University of Rome ``La Sapienza''. Data sets on the Flora of Lazio and Flora of Rome have been 
transcribed respectively from Anzalone \cite{Anzalone1984} and Celesti Grapow 
et al \cite{CelestiGrapow1995}. 
For data sets on the Flora of the Coliseum in Rome through centuries we referred to 
Caneva et al. \cite{Canevaetal2002}.  

In all data sets we consider a taxonomic classification tree made up of nine levels: 
species, genus, family, 
order, subclass, class, sub-phylum, phylum and sub-kingdom. All species are classified according to the 
nomenclature by A. Cronquist \cite{Cronquist1998,Jeffry1982,Bridsonetal1992,Weieretal1992}. 
We adopt a graph representation of the taxonomic tree. We assign a vertex for each taxon and an 
edge is drawn 
between two vertices $i,j$ if the corresponding subtaxon $j$ 
belongs to the taxon $i$. At the highest level, all 
species are eventually grouped in a same taxon; this implies that the resulting 
topology is that of a particular type of graph called a  ``tree'', i.e. a 
connected graph without loops in which any vertex can 
be reached starting from a specific node named root or source. 

\subsection{Random Subsets}
To compare the results obtained for the data sets with a null hypothesis we consider 
random species collections. We start from a pool of about $100,000$ species 
from all around the world (a fraction 
of all vascular plants in the world), that contains all the 
species in the considered data sets. Each data set represents a subset of the pool corresponding to species 
collections with specific 
geographic and climatic properties. From this pool we extract different random collections of 
species with sizes comparable to those of the real species assemblages. If we want to generate a 
randomized version of 
an ecosystem of $100$ species we simply select $100$ species 
from the entire pool of $100000$ species with uniform probability. Note that 
only the taxa corresponding to the species level are included in the pool. The random 
extraction is carried on only at the lower level of the taxonomic tree. 
The entire taxonomic classification 
from the genus up to the phylum is known for each element in the pool. 
The taxonomic tree corresponding to 
the random collection is thus recreated by clustering together the extracted 
species that happen to belong 
to the same genus and by iterating this clustering procedure until we have reconstructed 
the nine levels taxonomic 
tree of every random collection.  We then consider the frequency distribution of 
the degree of these trees. 
 
\section{Results} 
For every data set we represented the entire taxonomic classification as a nine level tree graph 
(see Data and Methods for details). 
We computed the frequency 
distribution of the degree, defined as the probability to have $k$ subtaxa for 
every taxon in the all tree, regardless the specific taxonomic level to which 
the different taxa actually belong. We observed a universal broad 
distribution, independent upon the geographic and climatic features of the different 
species assemblages 
(Fig. 1). 
\begin{figure}
\centerline{
\includegraphics{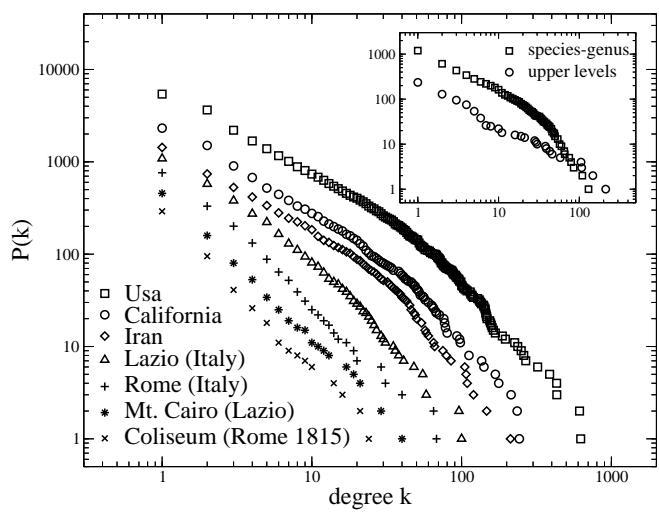}
}
\caption{Frequency distribution of the degree for some of the considered species 
collections (we show here the cumulative 
distribution which represents the distribution of the number of taxa containing 
more than k subtaxa). The distribution is 
universal through different assemblages. We observe a clear bending around $k=20$ which becomes more evident 
as the size of the system increases from 
about $250$ (bottom) to about $33000$ (top). Since higher taxa definition is 
somewhat arbitrary we show what is the 
effect of them on the statistics of the whole tree. In the inset, we make a 
comparison between the frequency distribution 
of the degree for the entire tree of the Flora of Iran (black diamonds) and 
the contribution of the species-genus levels 
(red diamonds) and of the rest of the tree (green diamonds) from the family up 
to the phylum of the same data set }
\label{Fig1}
\end{figure}

In all data sets, the initial slope of the frequency distribution (small values of the degree $1 < k < 20$) is in the
range $[1.9-2.7]$ and it decreases with the size of the system 
(number of species in the data set). For $k>20$ 
the distribution shows a clear bending that becomes more and more evident 
as the size of the system increases. We are not interested here in fitting the taxa distribution with any specific
function. As we will see in the following, we rather want to show that the properties of this distribution are
different if we consider random species collections instead of real florae. 
Interestingly, we found that
this result is robust in time. An example of this robustness is given by the analysis of six species collections 
representing the flora inside the Coliseum in Rome at six different 
historical periods from year $1643$ to 
$2001$ \cite{Canevaetal2002}. We observed that the behavior of the 
frequency distribution of the degree holds 
through temporal evolution of the system. 

To gain insight in the behavior of the frequency distribution of the degree and 
in the presence of the bending, 
we split the taxonomic tree into two parts. We compared the contributions given separately to the frequency 
distribution of the entire tree by the nodes belonging to the genus level and those belonging to all other 
levels. Fig. 1 (inset) shows the difference between the frequency distribution of the number of 
genera containing 
k species and that of the number of all upper taxa, from the family level up to the phylum, containing k subtaxa.
The first one is mainly responsible for the shape of the final distribution at small values of the degree and shows 
a very strong bending for $k>20$. The contribution of the second one is 
relevant for the smoothing of the bending at larger values of the degree.

Another remarkable universal feature across taxonomic levels emerges form the study of the topology. 
If we label each taxonomic level by $L$, we can define $n_L$ as the number of taxa at level $L$.
In order to understand the organization of the taxonomic levels, we defined $n_g=n_2$, $n_f=n_3$, 
and $n_o=n_4$, 
respectively as the 
number of genera ($L=2$), families ($L=3$) and orders ($L=4$) in our species assemblages, and 
we measured the relationship between these 
numbers and the number $n_s=n_1$ of species ($L=1$) in each data set. 
The resulting distributions are well fitted by power-law functions with 
exponents less than 1 (Fig. 2). As expected \cite{Enquistetal2001}, 
the number of higher taxa rises at a slower rate than species richness.
But what is more surprising is that for these taxonomic levels the number $n_{L+1}$ versus $n_L$ 
scales as a power-law 
\begin{equation}
n_{L+1}\propto n_L^\alpha
\end{equation}
 with universal exponent $\alpha =0.61$ independent upon $L$. 
This property can be understood from the main panel of Fig. 2 by 
noting that $n_2 \propto n_1^\alpha$ with $\alpha=0.61$ is the scaling between $n_g$ and $n_s$, 
while $n_3 \propto n_2^\alpha \propto (n_1^\alpha)^\alpha \propto n_1^\beta$ 
 with $\beta=\alpha^2=0.37$ is the scaling between $n_f$ and $n_s$, 
and finally $n_4 \propto n_3^\alpha \propto ((n_1^\alpha)^\alpha)^\alpha \propto n_1^\gamma $  
with $\gamma=0.23$  (very close to the fitted value 0.24) is the scaling between $n_o$ and $n_s$. 
This remarkable universality, on which we comment later, has never been documented before 
in the literature to the best of our knowledge.

\begin{figure}
\centerline{\includegraphics{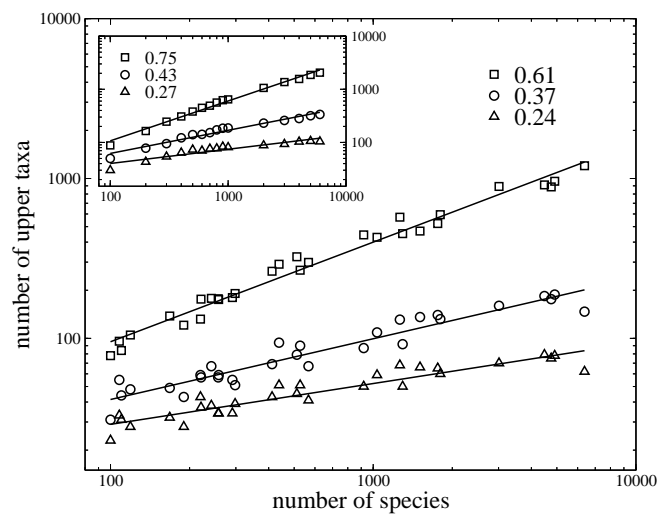}}
\caption{
Relationship between the number $n_g$ (squares), $n_f$ (circles), and $n_o$ (triangles), 
respectively the number of genera, families and orders in our species assemblages, and the number of 
species $n_s$. The resulting distributions are well fitted by power law functions with exponents less 
than $1$ showing that the number of higher taxa rises at a slower rate than species 
richness \protect \cite{Enquistetal2001}.
Moreover, the number $n_{L+1}$ of taxa at level $L+1$ versus the number $n_L$ of taxa at level $L$ 
is as a power-law with universal exponent $\alpha=0.61$. 
In the inset the same relationships in the case of random collections of species (results obtained 
as an average over $100$ random collections for each value of $n_s$). The three exponents are 
greater than in the real case suggesting that real species assemblages show a greater 
taxonomic similarity than random ones. 
Moreover, in this case the scaling exponent between $n_{L+1}$  and  $n_L$ is not universal.  
}
\label{Fig2}
\end{figure}
 
We compared our florae to a null model of random collections of species 
in order to test and quantify the 
significance of these universal results. We considered a pool of about 
$100,000$ species from all around 
the world (containing also all the species in the considered 
data sets) and we randomly extracted 
different collections with sizes comparable to those of the real species 
assemblages. We reconstructed the 
taxonomic tree of these random collections and studied the frequency 
distribution of the degree 
(see Data and Methods for details).  In Fig. 3 we compare the degree distribution of a 
real species assemblage to the one corresponding to a random collection of the same size.

\begin{figure}
\centerline{\includegraphics{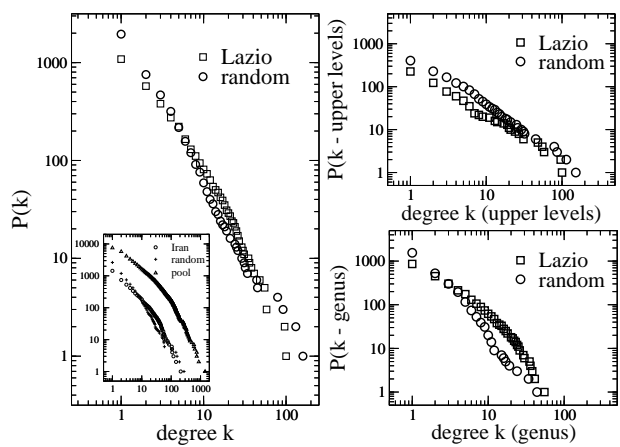}}
\caption{Frequency distribution of the degree for the Flora of Lazio and for a random 
collection of species of the same 
size about ($3000$ species). In the inset, the same comparison in the case of the Flora of Iran; 
the frequency distribution of the entire pool of $100,000$ is also shown. 
On the right the comparison between the specific distributions in the case of the species-genus levels 
(bottom) and of the rest of the tree (top). The random data are always obtained as a mean over 
$100$ realizations (random sampling of the big pool of species).}
\label{Fig3}
\end{figure}

Finally, we studied the frequency distribution of the number $n_s(g)$ of species in each genus $g$, 
normalized by its average $\langle n(g) \rangle$ and we compared results for a species 
assemblage and a random 
collection of species. Starting again from our pool of about $100,000$ species, we considered a large 
number of different random distributions of a number of species, corresponding to the size of the real list studied, 
among the different genera present in the pool. At the end we obtained a mean occupation 
number $\langle n(g) \rangle$ for each genus. We then considered a real species assemblage 
and a single random collection of species of the same size and we measured the degree $n_s(g)$ of each genus in both 
cases. We computed the frequency distribution of the ratio 
$n_s(g)/\langle n(g) \rangle$ and we compared a real species distribution of species 
among genera with a random one. 
The two distributions are shown in Fig. 4 (with a comparison between the Flora of 
Lazio and a random collection of the same size; similar results hold for every real species assemblage). 

\begin{figure}
\centerline{\includegraphics{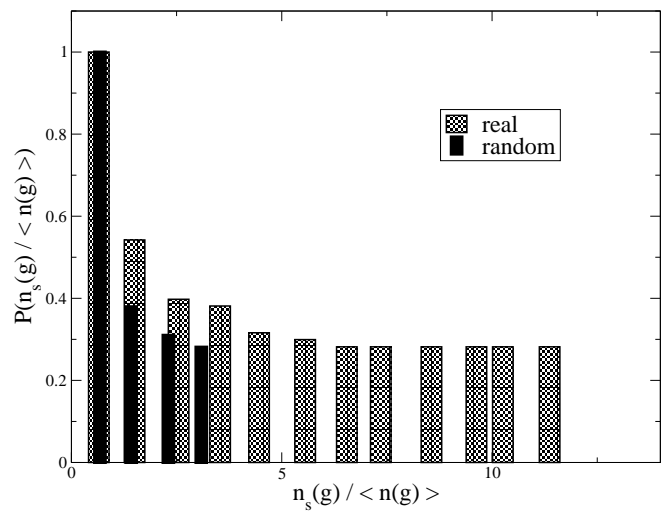}}
\caption{Distribution of the normalized number of species in each genus for a 
random uniform process and for a real flora 
(Flora of Lazio, about $3000$ species). The deviations observed in real flora are 
one order of magnitude larger than the standard deviation (here equal to $0.54$) 
that indicates the existence of an ``attraction effect'' 
between species leading to a smaller genus diversity. 
The random data are obtained by averaging over 100 realizations.}
\label{Fig4}
\end{figure}

\section{Discussion}
Recently, a great variety of social, technological and biological systems have 
been represented as graphs. In most 
cases the statistical properties of such graphs are universal. The frequency 
distribution of the number of links 
per site k (i.e. the degree) is distributed according to a power-law of the kind 
$P(k) \propto k^{-\gamma}$ with an exponent $\gamma$ between $2$ and $3$. 
This result suggests the possible existence of a 
unique mechanism for the onset of such common features \cite{Albertetal2002}. 
In the case of real species assemblages the 
frequency distribution of the degree of the corresponding taxonomic tree 
cannot be properly fitted by a power-law but still
the functional form appears to be robust across geography and time. 
Instead of trying to find the proper 
fit for such quantity, we concentrate on the different properties we can 
measure for real florae and random 
collections. 
As we will see in detail, the deviations from a power-law function actually 
contain the relevant information on the presence of a correlation 
between different species populating the same area. 
Fig. 1 shows the frequency distribution of the degree of the taxonomic tree 
for some of the florae considered in this 
study. The result is stable in time and 
independent of the geographic and climatic properties of the specific data set. The 
characteristic shape of the distribution 
derives from a different behavior of the branching process across different 
levels of the taxonomic tree. The 
comparison of the degree distribution at different taxonomic levels suggests 
that the taxonomic tree is far from 
being equilibrated. The origin of the bending in the distribution of the entire 
tree seems to lie in the distribution 
of the degree at the genus level, which strongly contributes to the smaller 
values of the degree, while the distribution 
of the upper levels contributes to the tail of the distribution lightening the 
bending. Moreover, the study of the number 
of higher taxa as a function of the species richness (Fig. 2) reveals that the 
number of species increases more rapidly 
than the number of taxa at the upper levels. This behavior propagates through 
the taxonomic tree: the number of families 
increases at a slower rate compared to the number of genera but it rises faster 
than the number of orders. 
More precisely, the number of taxa at a given level versus the number of taxa at the 
lower level is a universal power law with exponent $\alpha=0.61$ which is level-independent. 
Therefore, even if the details of the branching process differ from one level to the other, 
the overall taxonomic diversity measured by the number of taxa at a given level obeys a 
surprisingly regular scaling from the leaves to the root. This suggests that the observed 
scaling behavior reflects some intrinsic organization principle underlying the taxonomic 
structure of real ecosystems. This result is confirmed by the analysis of randomized 
data (see below) where this universality is destroyed. 

We also measured the mean-degree of the genera for different sizes of the data sets and 
we found that it increases with size. 
In bigger assemblages the number of genera obviously increases but with an exponent 
smaller than $1$, meaning that a 
greater number of species is distributed among a proportionally smaller number of 
taxa. The combination of these two 
results, together with the fact noticed previously that different taxonomic levels 
contribute to different regions of 
the degree distribution, could explain why the frequency distribution of the degree of the entire 
tree does not change in the same way for 
different values of k while increasing the size of the system. Even if taxonomic 
classification is somewhat arbitrary, 
these results indicate a precise trend. There are examples of 
branching process models that try to 
reproduce similar behaviors in the distribution of subtaxa per taxon at specific 
taxonomic levels 
\cite{Yule1924,Epsteinetal1989,Chuetal1999}.
This behavior of the degree distribution of taxonomic trees either reflects a 
topological property of the general 
structure of the taxonomic classification (i.e. the taxonomic tree of all the 
plants in the world follows this peculiar 
distribution and so does every subset) or it reflects the existence of some 
interaction between different species in the 
same environment.  We now discuss evidences that the latter hypothesis holds. 

The most immediate test of the significance of the results presented above is a 
comparison between observed data sets 
and a null model of random collections of species. The frequency distribution of 
the degree in the real and random cases 
is shown in Fig. 3. If we consider the entire tree, we find that the initial slope of the distribution (small
values of the degree $k<10$) is different in the two cases. In Fig. 3 (left) we have $2.1$ for the Flora 
of Lazio 
and $2.5$ in the case of a random collection of the same size extracted 
from the entire pool (similar 
results with specific values of the slope hold for any of the observed data sets). 
Moreover, for $k>10$ the bending, that is 
evident in the case of the real species assemblage, is not present in the random 
case. In the inset, we present the same 
comparison for the Flora of Iran. The third distribution given in the plot 
corresponds to the entire pool 
of about $100,000$ species. Note that, unlike the random collection, the real 
system is able to reproduce the behavior 
of the pool. Once again we consider separately the distribution at the 
species-genus level and for the rest of the tree 
(Fig. 3 right). In the first case we have a different slope in the first part of 
the distribution and very soon a divergence 
of the two distributions (Fig. 3 right-bottom). In the second case the 
distribution corresponding to the random collection 
of species overestimates the one relative to the observed data at all values 
of the degree (Fig. 3 right-top). Since the 
two data sets are subsets of a same pool of species and are classified according 
to the same taxonomic structure, these results provide a quantitative measure 
of the expected long-range correlations between different 
species in the same environment. These correlations shape the taxonomic structure 
of real species collections generating 
specific features that cannot be shared by a random subset. Differences are 
actually not very strong and one could object 
that real data sets are simply biased random collections. But fundamental 
ecological assumptions together with previous 
and further considerations in our discussion strengthen the idea that Nature 
actually follows a non-random path in the 
organization of an ecosystem and that differences between real and random 
collections, even if slight, are to be 
considered as an evidence of this non-random behavior. Moreover we have observed 
that an increase in the size of the 
big pool of species from which we extract the random collections results in stronger 
differences between the real 
data sets and the randomized one.

An even more convincing confirmation of the presence of correlations between species in 
the same environment comes from the 
comparison of the behavior of the number of superior taxa versus the number of 
species. In the 
random case the variation in the number of genera, families and orders with respect to the 
number of species is 
qualitatively similar than in real systems, but the values of the exponents are always greater 
(Fig. 2 inset). This result points out that species in the same environment 
are more taxonomically 
similar then random collections of species. From an ecological point of view, 
co-occurring species that experience similar environmental conditions 
are likely to be more taxonomically similar than ecologically distant species, because of the process 
of environmental filtering. During evolutionary diversification species traits tend to be preserved 
within a taxon. As a consequence of this, the species capabilities of
colonizing a given piece of physical space with a certain set of environmental conditions is thought to depend to 
some degree on their taxonomic similarity \cite{Webb2000}.
In the random case the universality of the scaling exponent relating the number of 
taxa at a given level with that of taxa at the lower level is destroyed. This is easily seen 
because, if universality held in the random case too, the families-genera scaling exponent 
($0.43$) and the orders-families one ($0.24$) would be respectively the second and the third 
power of the genera-species scaling exponent ($0.75$) as in the case of real data. Instead we have $0.75^2\simeq 0.56>0.43$ 
and $0.75^3\simeq 0.42>0.27$. Therefore the universality displayed by real data is not 
accidental and appears to be a meaningful characteristic of co-evolved communities of 
species which is not shared by random species assemblages.

Finally the mean degree of the genera grows 
more slowly with the number 
of species in the random case than in the real one, confirming the idea of a greater 
taxonomic diversity of random 
species collections. A wide range of evolutionary models suggests that this correlation 
between species could lie 
on evolutionary mechanism such as speciation, selection and competition \cite{Camachoetal2000}. 
One possible explanation 
of the presence of power-laws in taxonomy systems has been proposed in the framework of the spin glass 
theory \cite{Epsteinetal1989}. In the simplest version, a spin glass (considered as a 
paradigmatic model of disordered systems) 
consists of randomly coupled binary variables (spin up or spin down) defined for every node of a 
regular lattice. The idea 
is that phenotype traits can be coupled together in a similar fashion. Some aspects of 
evolution could then be 
described by the same mathematical framework as the one employed in the description of a spin-glass system. 
The latter is known to have equilibrium 
states characterized by a hierarchy of sub-equilibrium states. Ultimately, it is the 
intrinsic disorder (in this 
case the genetic codes of the various species) that naturally produces the observed scale 
invariance. This can be 
also empirically found by means of a cellular automaton model \cite{Caldarellietal2002}.  
It is interesting to note 
that ,as the exponent $\gamma$ of the degree distribution has a value between 
$2$ and $3$, the distribution itself has a finite average value and diverging fluctuations 
around this value. 
Biologically, this means that, even if the mean number of off-springs in a taxon can be 
defined, every now and 
then a taxon with an huge number of off-springs appears. This quantitative analysis 
suggests that 
taxonomic trees are related to the wide class of scale-free networks present in epidemics 
\cite{Bogunaetal2003} 
as well as in protein interactions \cite{Giotetal2003}.

We discuss one last result in agreement with our previous considerations. 
We studied the distribution of species 
among genera. The results given in Fig. 4 indicate that in real assemblages 
species are not randomly distributed 
among genera. The frequency distribution of the ratio $n_s(g)/\langle n(g)\rangle$ 
(degree of each genus normalized to the mean occupation number) decays very 
rapidly in the random case. The deviations from the observed distribution are one order of magnitude 
larger than the standard 
deviation, spotting that in real assemblages there is an effective attraction between species that leads to 
over-represented genera, while random species collections tend to spread 
among a greater number of 
genera (less species for each genus). This result is in good agreement 
with the previous assertion 
that real data sets are more taxonomically similar than random collections 
of species, and with the hypothesized 
existence of some universal mechanism leading species organization in nature.

In conclusion, our results confirm that a taxonomic tree is not simply 
an artificial classification of 
species, but it contains information on the processes that shape the corresponding 
flora. We have studied the 
frequency distribution of the degree for the taxonomic trees of several 
data sets corresponding to 
collections of species from different geographical and climatic environments 
and we have revealed the presence of universal stable properties. 
We have numerically characterized real species assemblages with respect to 
random collections 
spotting differences in the frequency distribution of the degree and confirming 
the idea that species 
in a same environment are characterized by greater taxonomic similarity than 
random ones. In particular, 
we have observed that genera diversity in real assemblages is usually (very) low. All 
these results suggest the 
existence of some long-range correlation between species in a same environment 
and they could help in the 
realization of a new model able to reproduce such properties. Finally, we conclude with a cautionary remark: 
in this paper we used traditional Linnaean classification and not phylogenetic data, although the latter may seem
preferable. In principle, the species taxonomic information could easily be derived in a full phylogenetic context, 
by replacing Linnaean taxonomic trees by proper phylogenetic trees. However, a major obstacle in using a 
phylogenetic approach is that there are no phylogenies available for all plant species being studied and that
phylogenization of most interesting species is unlikely to occur within the next several years.
Moreover, we encounter two main limitations 
in applying a topological approach to phylogenetic trees[12]. 
First of all, the structure of the phylogenetic tree is unstable and it depends on the 
species included in such a way that the
number of nodes between two taxa is determined by the particular reference 
phylogeny used for the classification. As a result, the resulting
measures can only be used to compare assemblages whose species are a subset of the species in the 
reference phylogeny. Secondly, the species richness of a given taxon will influence its branching structure in the sense that two
species drawn at random from a species-rich taxon are likely to appear more topologically distant than two 
species from a species-poor taxon. For all these reasons, it is more realistic to remain within a Linnaean
taxonomic framework. We think that the results presented in this paper could 
thus contribute to a better 
understanding of the mechanisms ruling organization and biodiversity in plants 
ecosystems.

\ack{We acknowledge EU Fet Open Project COSIN IST-2001-33555 and European 
Integrated Project DELIS.}

\end{document}